\documentclass[showpacs,amsmath,amssymb,aps,showkeys,floatfix,prd,a4paper]{revtex4}

\usepackage[dvips]{graphicx}
\usepackage{dcolumn}
\usepackage{bm}
\usepackage{epsfig}
\usepackage{amsfonts}
\usepackage{amssymb,amscd}

\def\lsim{\raise0.3ex\hbox{$<$\kern-0.75em\raise-1.1ex\hbox{$\sim$}}}

\def\gsim{\raise0.3ex\hbox{$>$\kern-0.75em\raise-1.1ex\hbox{$\sim$}}}

\newcommand{\be}{\begin{equation}}

\newcommand{\ee}{\end{equation}}

\def\beq{\begin{equation}}

\def\eeq{\end{equation}}

\def\beqa{\begin{eqnarray}}

\def\eeqa{\end{eqnarray}}

\newcommand{\ba}{\begin{eqnarray}}

\def\gappeq{\mathrel{\rlap {\raise.5ex\hbox{$>$}}

{\lower.5ex\hbox{$\sim$}}}}

\def\lappeq{\mathrel{\rlap{\raise.5ex\hbox{$<$}}

{\lower.5ex\hbox{$\sim$}}}}

\def\Toprel#1\over#2{\mathrel{\mathop{#2}\limits^{#1}}}

\begin{document}

\begin{flushright}
LU TP 15-55\\
December 2015
\vskip1cm
\end{flushright}

\title{Phenomenological implications of the intrinsic charm in the $Z$ boson production at the LHC}
\author{G. Bailas $^{1,3}$ and  V. P. Gon\c{c}alves$^{2,3}$}
\affiliation{$^1$ Laboratoire de Physique Corpusculaire, Universit\' e Blaise Pascal, CNRS/IN2P3, 63177, Aubi\`ere Cedex, France \\
$^2$ Department of Astronomy and Theoretical Physics, Lund University, SE-223 62 Lund, Sweden \\  
$^{3}$ High and Medium Energy Group, Instituto de F\'{\i}sica e Matem\'atica,  Universidade Federal de Pelotas\\
Caixa Postal 354,  96010-900, Pelotas, RS, Brazil.
}

\begin{abstract}
In this paper we study the $Z$, $Z+$ jet, $Z+c$  and $Z+c+$ jet production in $pp$ collisions at the LHC considering  different models for an intrinsic charm content of the proton. We analyse the impact of the intrinsic charm in the rapidity and transverse momentum distributions for these different processes. Our results indicated that differently from the other processes, the $Z+c$ cross section is strongly affected by the presence of the intrinsic charm. Moreover, we propose the analysis of the ratios $R(Z+c/Z) \equiv \sigma(Z+c)/\sigma(Z)$ and   $R(Z+c/Z+\mbox{jet}) \equiv \sigma(Z+c)/\sigma(Z+\mbox{jet})$ and demonstrate that these observables can be used as a probe of the intrinsic charm.
\end{abstract}



\maketitle

\vspace{1cm}

A complete knowledge of the partonic structure of the hadrons is fundamental to make predictions for {the Standard Model and to beyond Standard Model processes at hadron colliders}.  Since the early days of the parton model and of the first deep inelastic scattering  (DIS)  experiments, determining the precise form of the quark and gluon distributions of the nucleon has been a major goal of high energy hadron physics. Over the last 40 years {huge} progress has been achieved. In particular, data from HERA have dramatically improved our knowledge about the small-$x$ behaviour of the parton distributions functions (PDF's) \cite{paul}. Another important improvement has occurred in our knowledge about the heavy quark contribution to the proton structure \cite{hq_hera}. In the last years several groups have proposed different schemes to determine these distributions considering that  the heavy quark component in PDF's can be perturbatively generated by gluon splitting (See e.g. \cite{tung}). This component is usually denoted  {\sl extrinsic} heavy quark  component. Moreover, the possibility of an {\sl intrinsic}   component  has been studied in detail and  included in the recent versions of the PDF 
parametrizations \cite{cteq,ct14,jr,int_bottom}. 
The hypothesis of intrinsic heavy quarks (IHQ) is a natural 
consequence of the quantum fluctuations inherent to Quantum Chromodynamics (QCD) and amounts to  
assuming the existence of a $Q\bar{Q}$ ($Q = c,b,t$) as a non perturbative component in the 
hadron wave function.  A comprehensive review of the main 
characteristics of the  IHQ models can be found in \cite{pumplin,peng,brodsky_review}.  In the model proposed by Brodsky, Hoyer, Peterson and Sakay (BHPS) \cite{bhps}, the creation of the 
$Q \overline{Q}$ pair was studied in detail (For a discussion of other models for the IHQ see Refs. \cite{navarra,hobbs}). It was assumed that the nucleon 
light cone wave function has  higher Fock states, the first one being 
$|q q q Q \overline{Q}>$. The probability of finding the nucleon in this 
configuration was given by the inverse of the squared invariant mass of the 
system. Because of the heavy quark mass, this probability as a function of the 
quark fractional momentum, $P(x)$, is very hard, as compared to the one obtained 
through the DGLAP evolution. Although this model predicts the $x$ dependence of the intrinsic components, its normalization should be constrained by fitting the experimental data and still is an open question. In particular, two recent global analysis of the PDF's \cite{ct14,jr}
have analysed the importance of an intrinsic heavy quark component to describe a wide range of hard scattering data and obtained distinct conclusions. Basically, the current situation is such that new and more precise data are necessary to probe the hypothesis of intrinsic heavy quarks \cite{brodsky_comment}.

The presence of an intrinsic heavy quark component is expected to directly modify the cross section of the processes which are initiated by heavy quarks and indirectly to other processes, since the presence of the IHQ modifies the contribution of the other partons due the momentum sum rule. 
One  of the most striking properties of an IHQ state, such as 
$|uudQ\bar{Q}\rangle$, is that the heavy constituents tend to carry the largest fraction of the 
momentum of the hadron. Consequently,  in contrast to heavy quarks produced through  usual perturbative QCD, which emerge  
with small longitudinal momentum,  the intrinsic quark component gives rise to heavy mesons  
with large fractional momenta relative to the beam particles. Therefore, 
the existence of an intrinsic component is expected to modify, for instance, the $x_F$ and rapidity 
distribution of charmed particles (See, e.g. \cite{ingelman,kkss,vicnav}). Moreover, it can also lead 
to Higgs production at high $x_F$ \cite{brod_higgs}. Many of these features have been 
discussed in the pioneering works on intrinsic heavy quark  \cite{bhps,vb,hal}. 
 Another promising observable to search the intrinsic heavy quark component is the associated production of a heavy quark with a gauge boson, which is strongly dependent on the heavy quark distribution. In recent years, several studies 
about the impact of an intrinsic heavy quark component on the proton wave function in the $\gamma + Q$ and $Z + Q$ cross sections were performed \cite{stavreva,lik_plb,iran,lik_prd} , with particular emphasis in the  case $Q = c$, since the probability of an intrinsic bottom is expected to be suppressed by a factor $(m_c^2/m_b^2)$ \cite{bhps}. Our goal in this paper is to complement these previous studies by the analysis of the phenomenological implications of the intrinsic charm in the $Z$ boson production and related processes at the LHC energies. In particular, we will analyse the impact of the intrinsic charm in the rapidity and transverse momentum dependencies of the $Z$, $Z+$ jet, $Z+c$  and $Z+c+$ jet cross sections at $\sqrt{s} = 7$ TeV. The cross sections will be computed considering the next - to - leading order corrections using the parton - level MC generator MCFM - version 6.8 \cite{mcfm} and using as input the different models present in  the CT14 parton distributions \cite{ct14}. It is important to emphasize that  the MCFM does not take into account the  mass corrections on the incoming heavy quark  lines, which makes our calculations  unreliable close to threshold. Such subject have been discussed in the literature in the last years and advances on the  treatment of the intrinsic heavy quarks in the perturbative computation of hard processes in the full kinematical range were  recently presented in Ref. \cite{forte}.

\begin{figure}
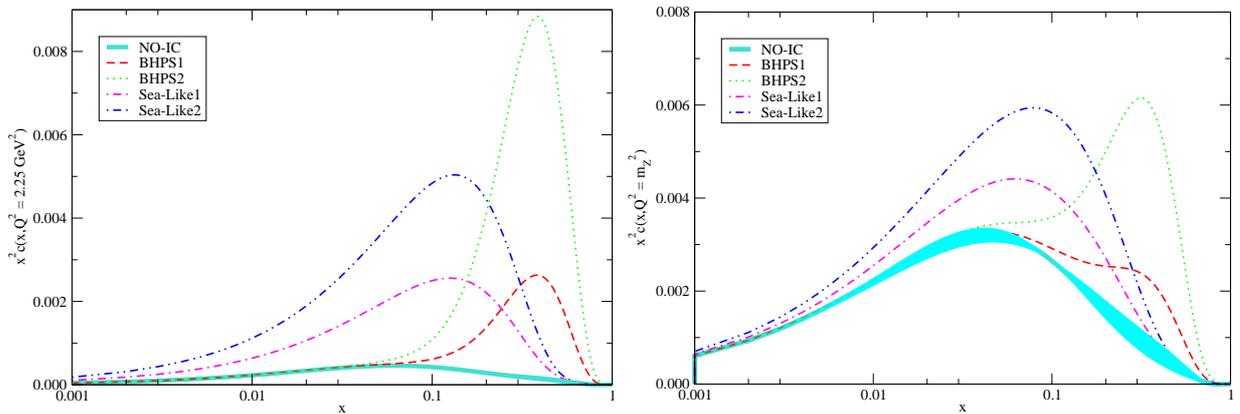

\begin{tabular}{cc}
{\psfig{figure=baixaenergia.eps,width=8cm}} &
{\psfig{figure=q2mz2.eps,width=8cm}}  
\end{tabular}
\caption{Comparison between the different models for the intrinsic charm  present in the CT14 parametrization for two different values of the hard scale $Q^2$.}
\label{fig1}
\end{figure}

{We will start} our analysis discussing the different models for the intrinsic charm (IC) present in the CT14 parametrization \cite{ct14}. As in the Ref. \cite{cteq}, the recent analysis performed in \cite{ct14} considered the intrinsic charm  as 
an ingredient in the global fit of DIS and hadronic colliders data and  determined the shape and normalization of the IC distribution in the same way as they do for other parton species. 
In fact they find several IC distributions which are compatible with the world data. {In addition to} the already mentioned BHPS model, the CTEQ - TEA group has tested another model of
intrinsic charm, called sea-like IC. It consists basically in assuming that at a very low resolution (before the DGLAP evolution) there is already some charm in the nucleon, which has a 
typical sea quark momentum distribution ($\simeq 1/ \sqrt{x}$) with normalization to be fixed by fitting data. In particular, they  provided four different sets of PDF´s which differ in the model for the IC and in the mean momentum fraction of the IC PDF,  $\langle x \rangle_{IC}$,  and are compatible with the current world data. These four sets will be used as input in our calculations and will be denoted by BHPS1 ($\langle x \rangle_{IC} = 0.6 \%$), BHPS2 ($\langle x \rangle_{IC} = 2.0 \%$), Sea -- like 1  ($\langle x \rangle_{IC} = 0.6 \%$).  and Sea -- like 2  ($\langle x \rangle_{IC} = 1.5 \%$). 
The different parametrizations are presented  in Fig. \ref{fig1} for two different values of the hard scale $Q^2$. For comparison we also present the NO-IC distribution, where the charm content of the nucleon sea comes from the DGLAP evolution (exclusive charm component). In this case we present the uncertainty band in this distribution obtained by the CTEQ - TEA group.   For $Q^2 = 2.25$ GeV$^2$ (left panel) the 
intrinsic charm distribution can be a factor ten (Sea -- like1) to twenty (BHPS2) larger than the NO-IC distribution  and the peaks of the IC distributions occurs in the large $x$ 
($\ge 0.1$) region. For $Q^2 = m_Z^2$ (right panel) the peaks decreases in magnitude and also shifts to smaller values of $x$. However, this peculiar behaviour still is present, which  gives us hope to observe IC experimentally in the charm - initiated processes.

\begin{figure}[!ht]
\begin{tabular}{cc}
{\psfig{figure=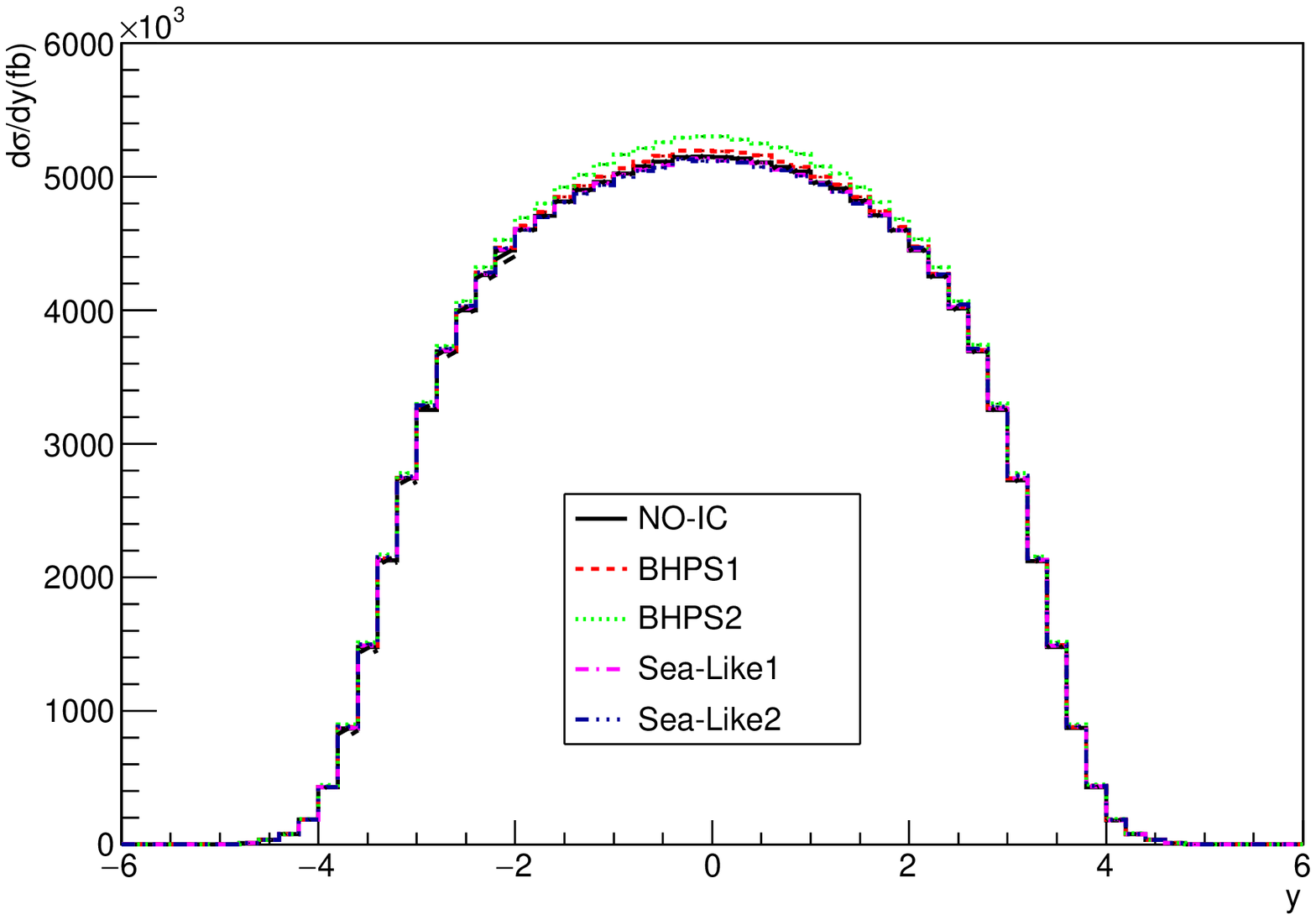,width=7cm}} & {\psfig{figure=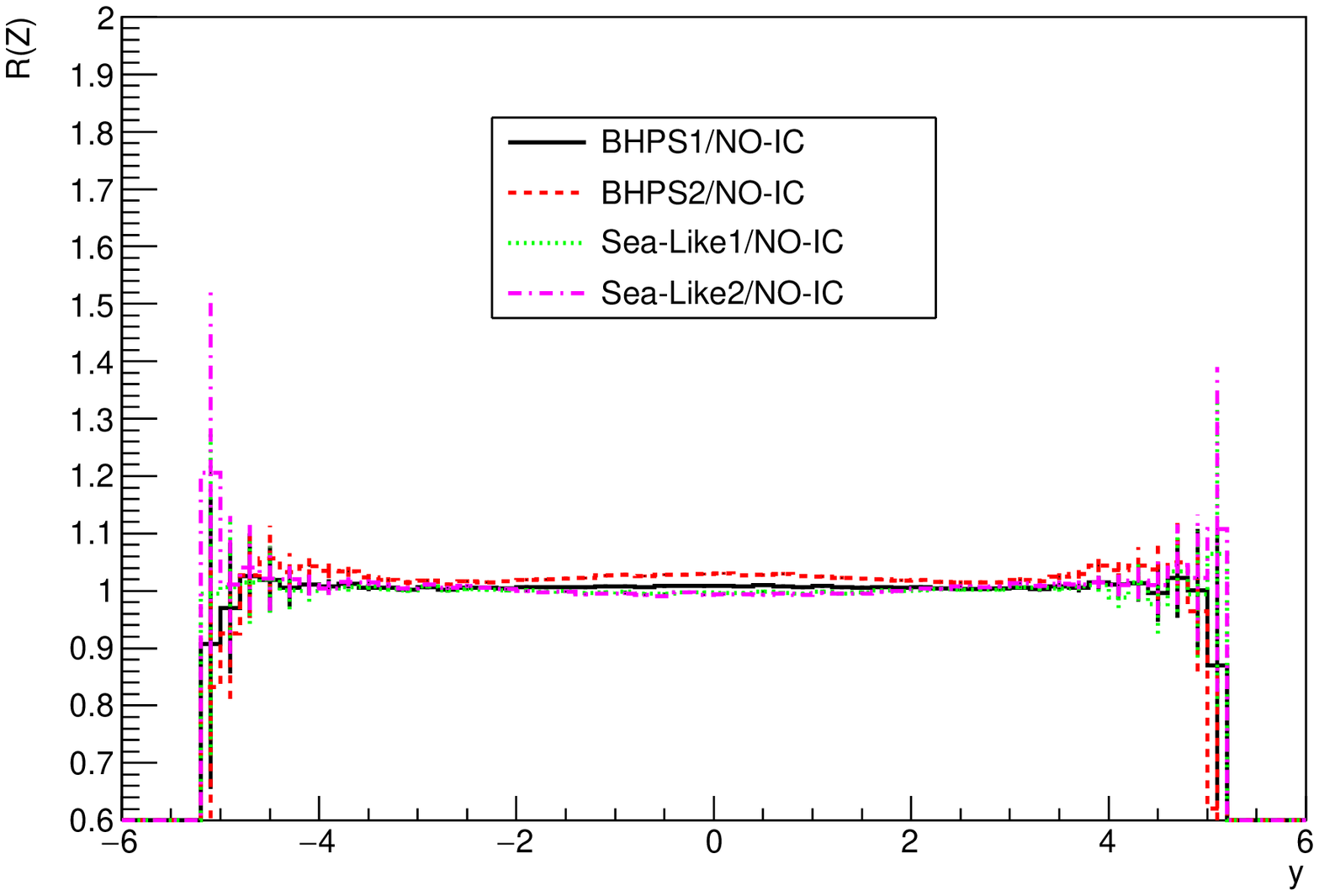,width=7cm}} \\
{\psfig{figure=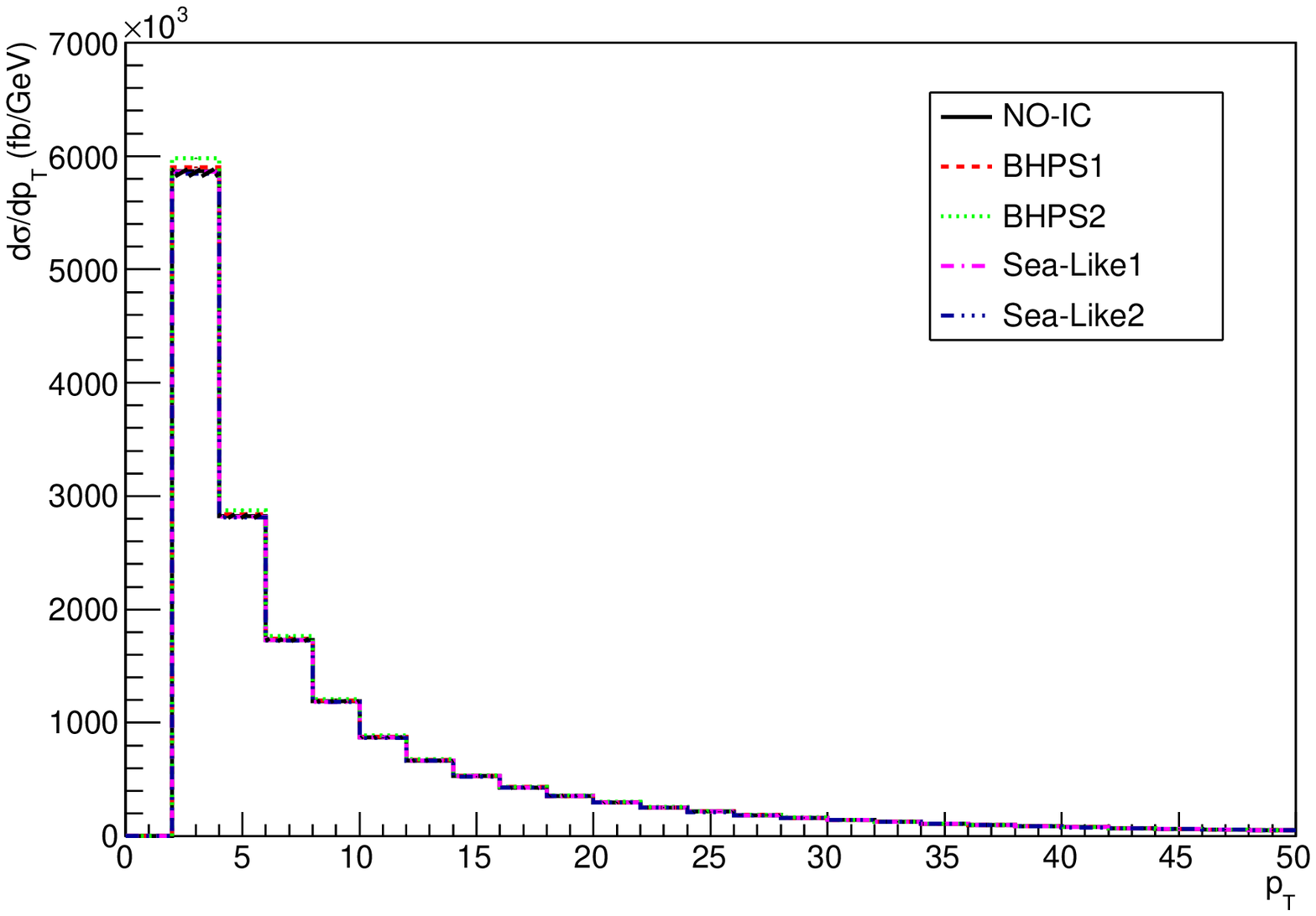,width=7cm}} & {\psfig{figure=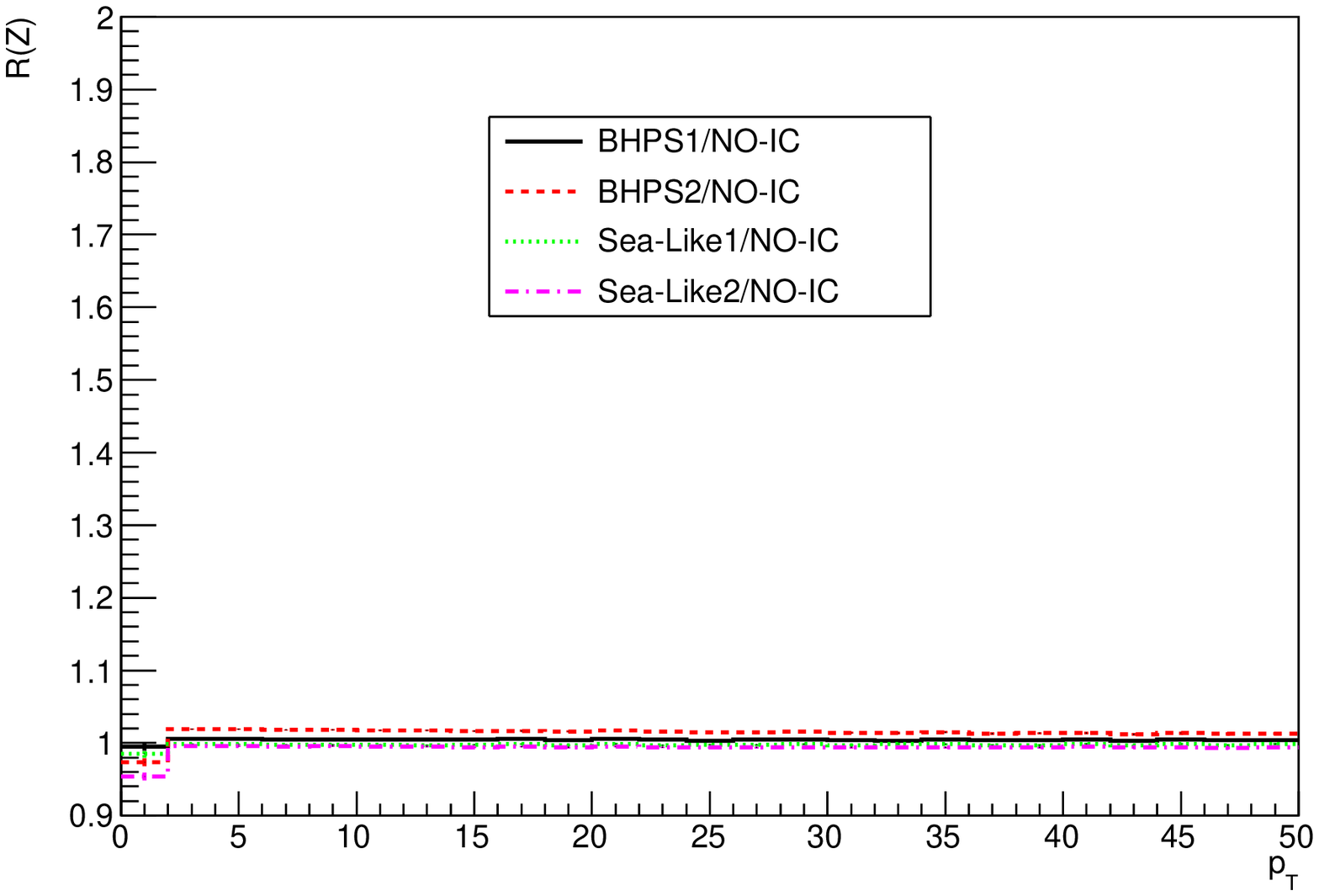,width=7cm}}
\end{tabular}                                                                                                                       
\caption{Left panels: Predictions for the rapidity and transverse momentum distributions for the $Z$ production  in $pp$ collisions at $\sqrt{s} = 7$ TeV. Right panels: Rapidity and transverse momentum dependencies of the ratio between the IC and central NO-IC predictions. }
\label{fig2}
\end{figure}

\begin{figure}[!ht]
\begin{tabular}{cc}
{\psfig{figure=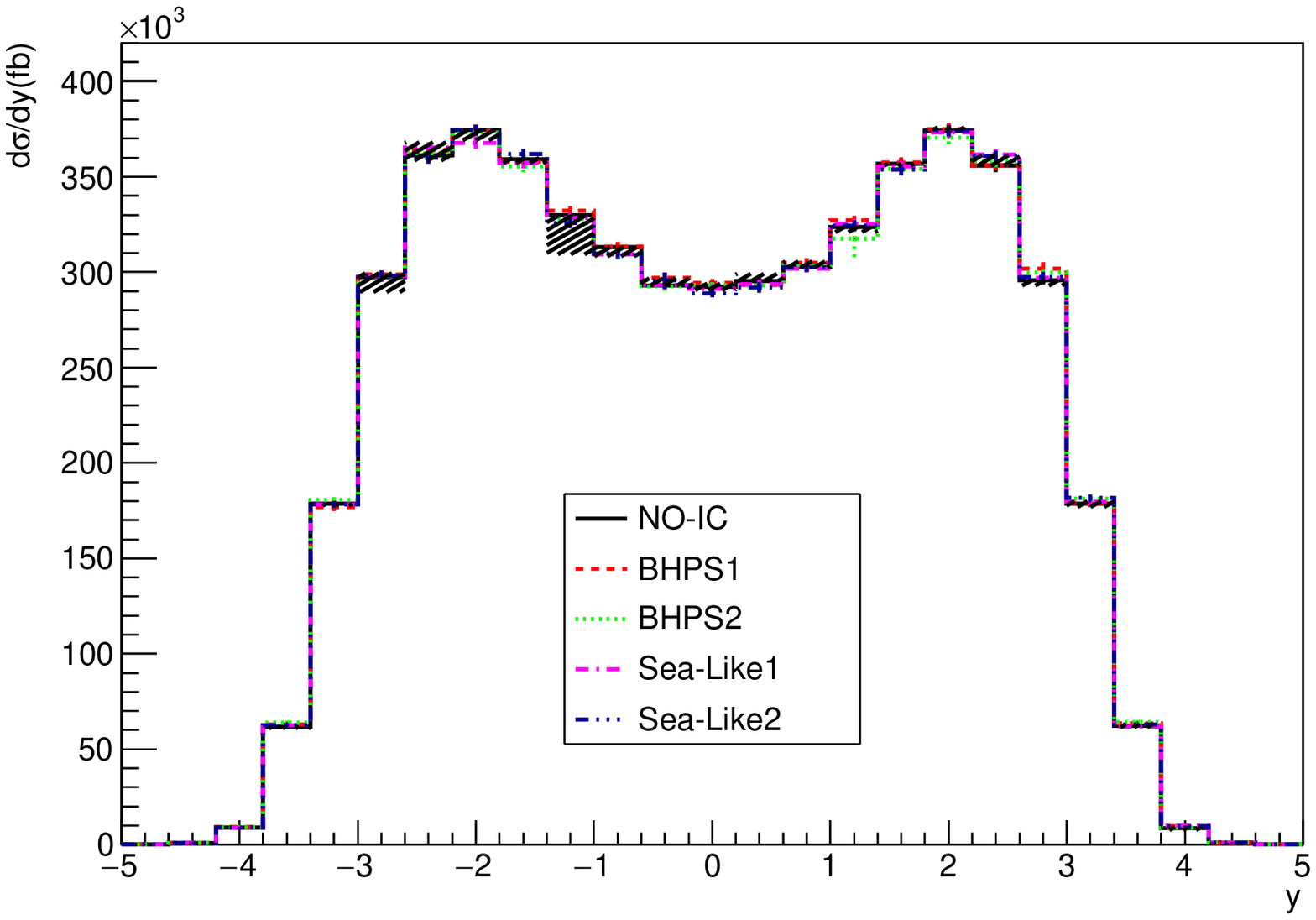,width=7cm}} & {\psfig{figure=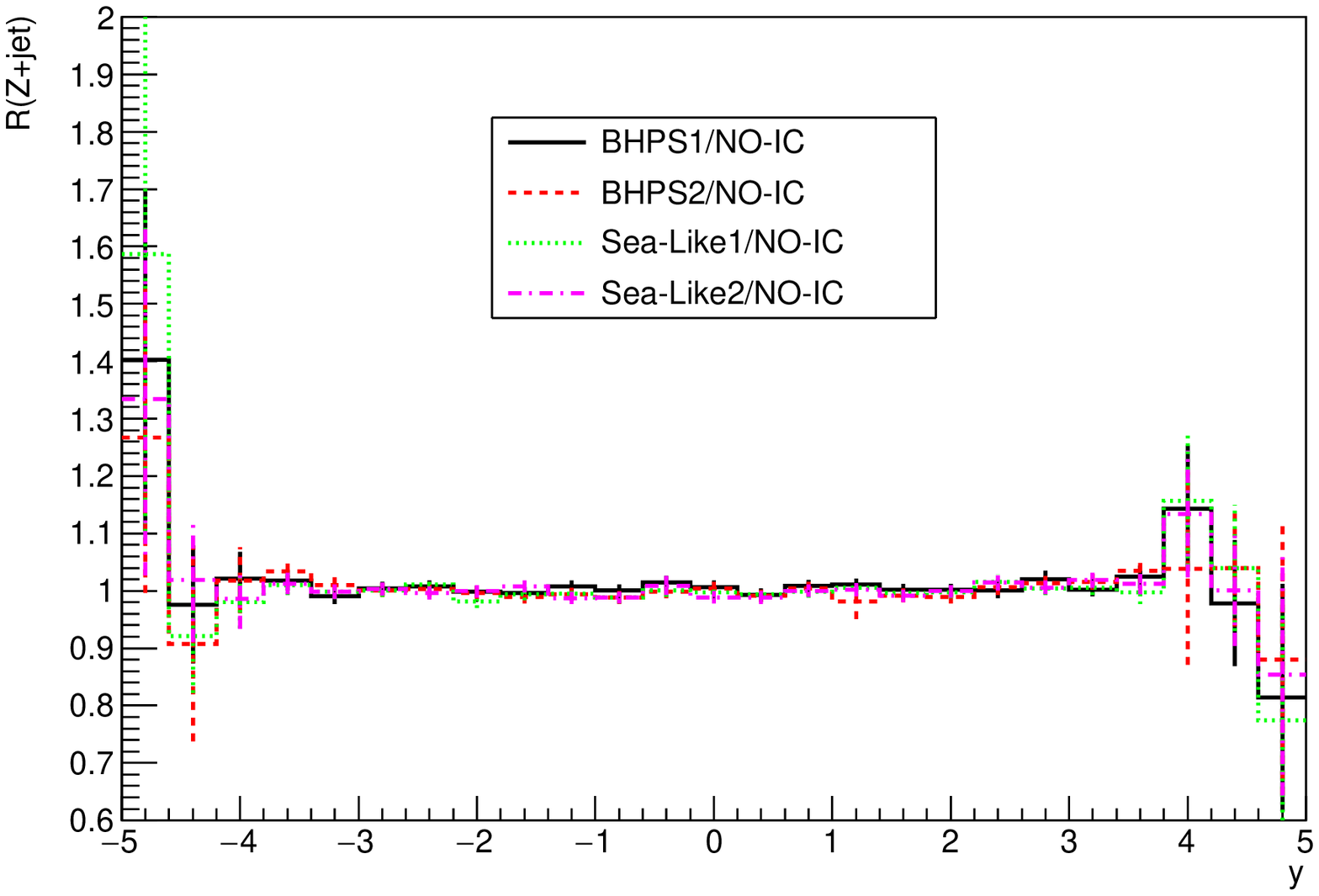,width=7cm}} \\
{\psfig{figure=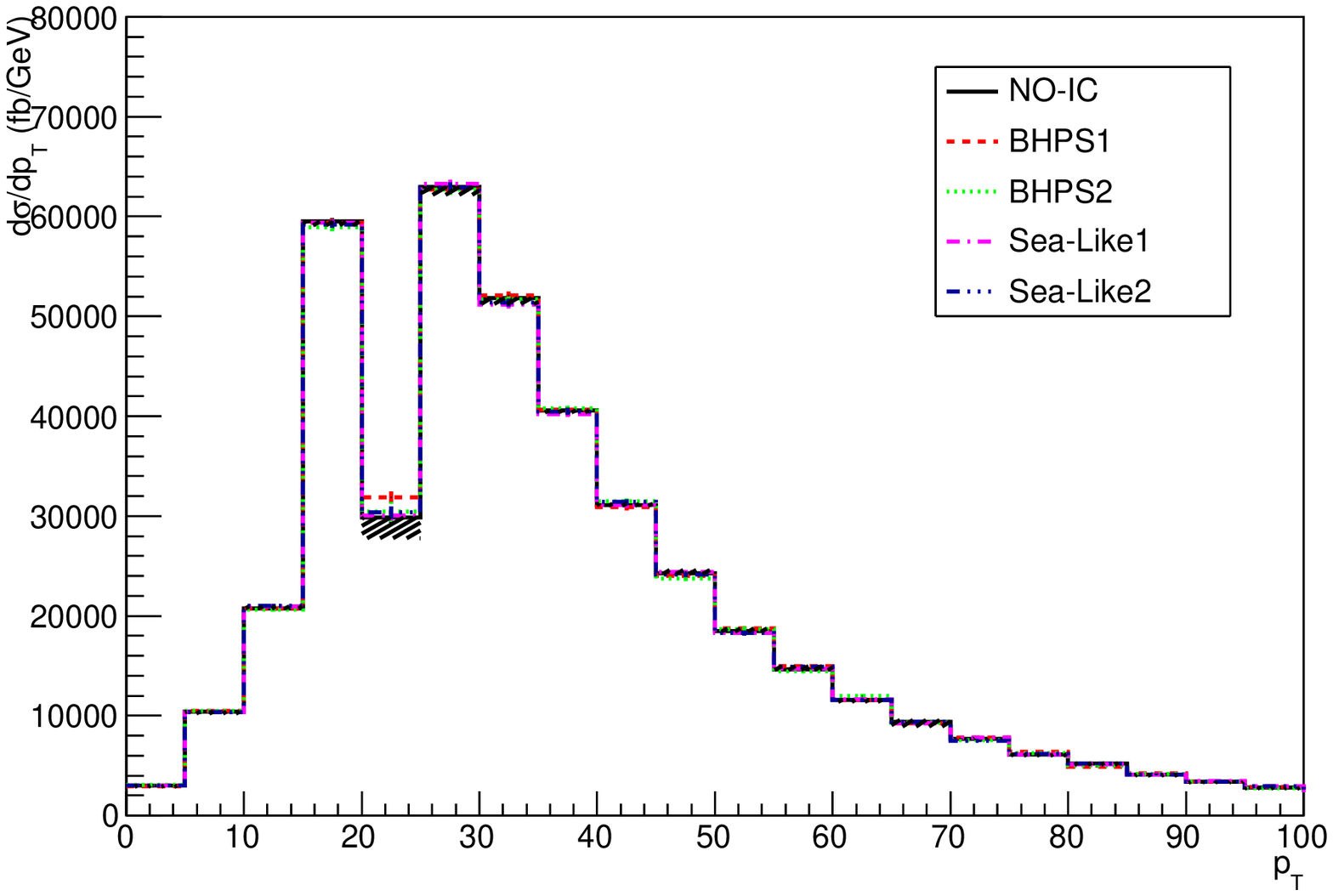,width=7cm}} & {\psfig{figure=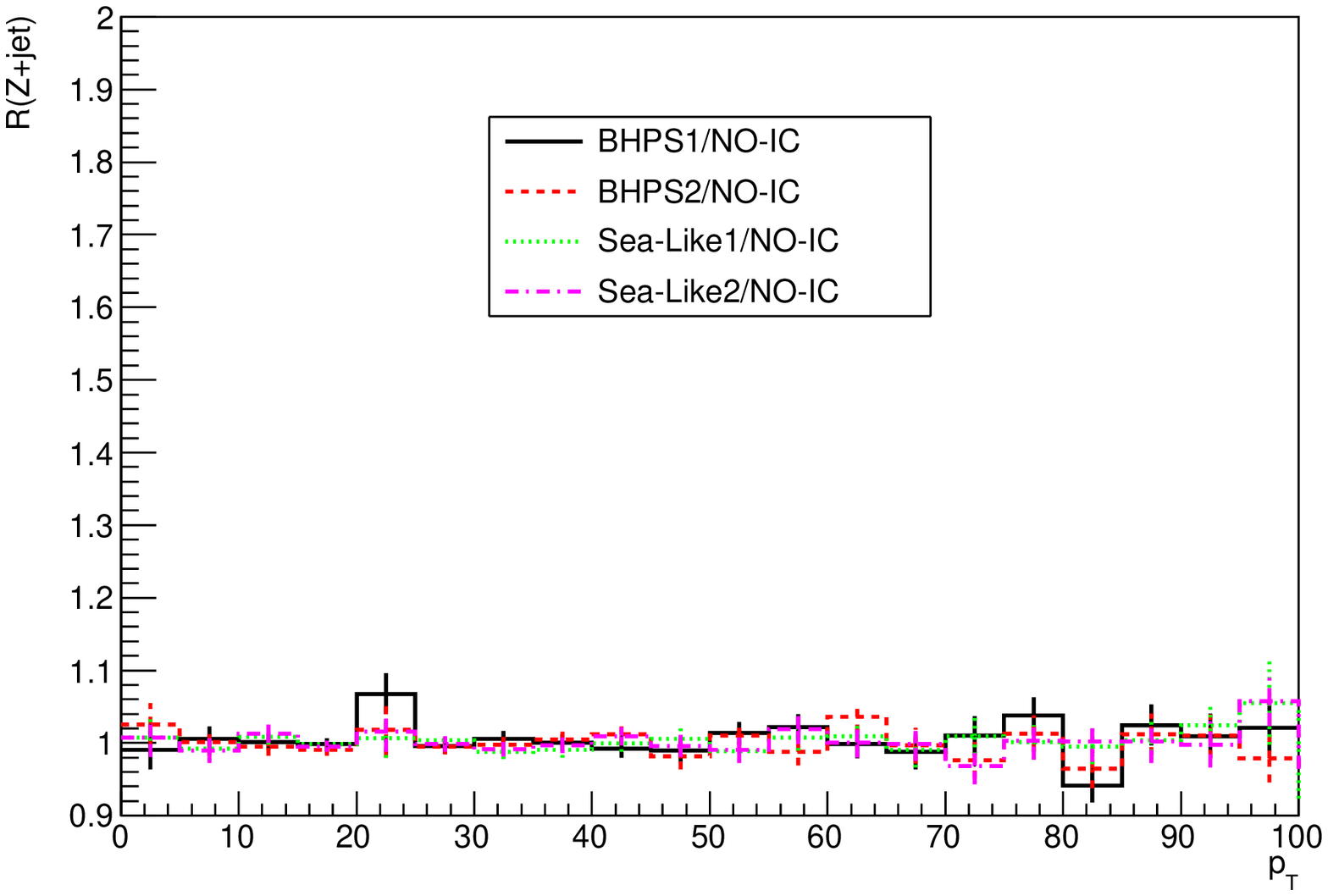,width=7cm}}
\end{tabular}                                                                                                                       
\caption{Left panels: Predictions for the rapidity and transverse momentum distributions for the $Z+$ jet production  in $pp$ collisions at $\sqrt{s} = 7$ TeV. Right panels: Rapidity and transverse momentum dependencies of the ratio between the IC and central NO-IC predictions.}
\label{fig3}
\end{figure}

\begin{figure}[t]
\begin{tabular}{cc}
{\psfig{figure=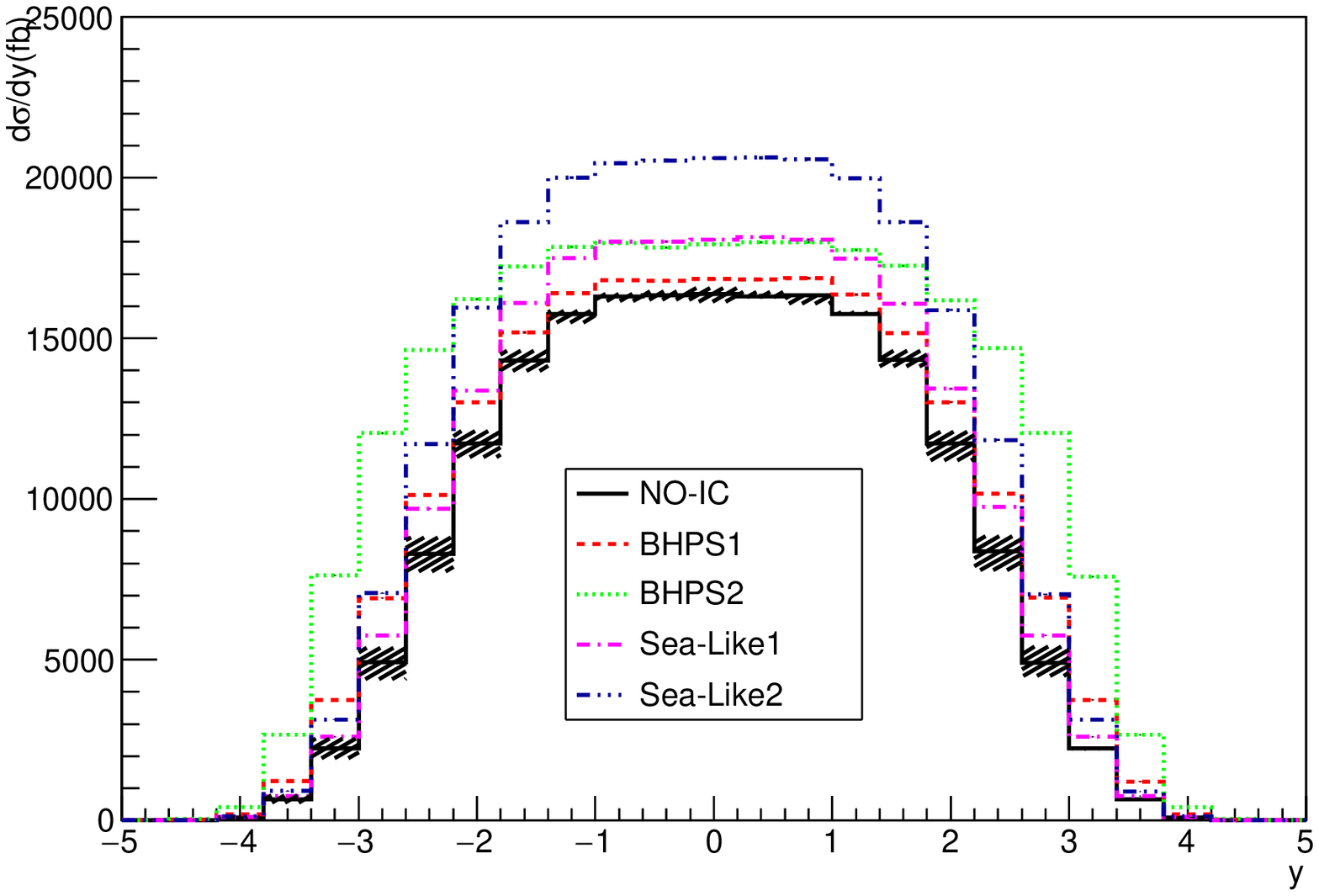,width=8cm}} & {\psfig{figure=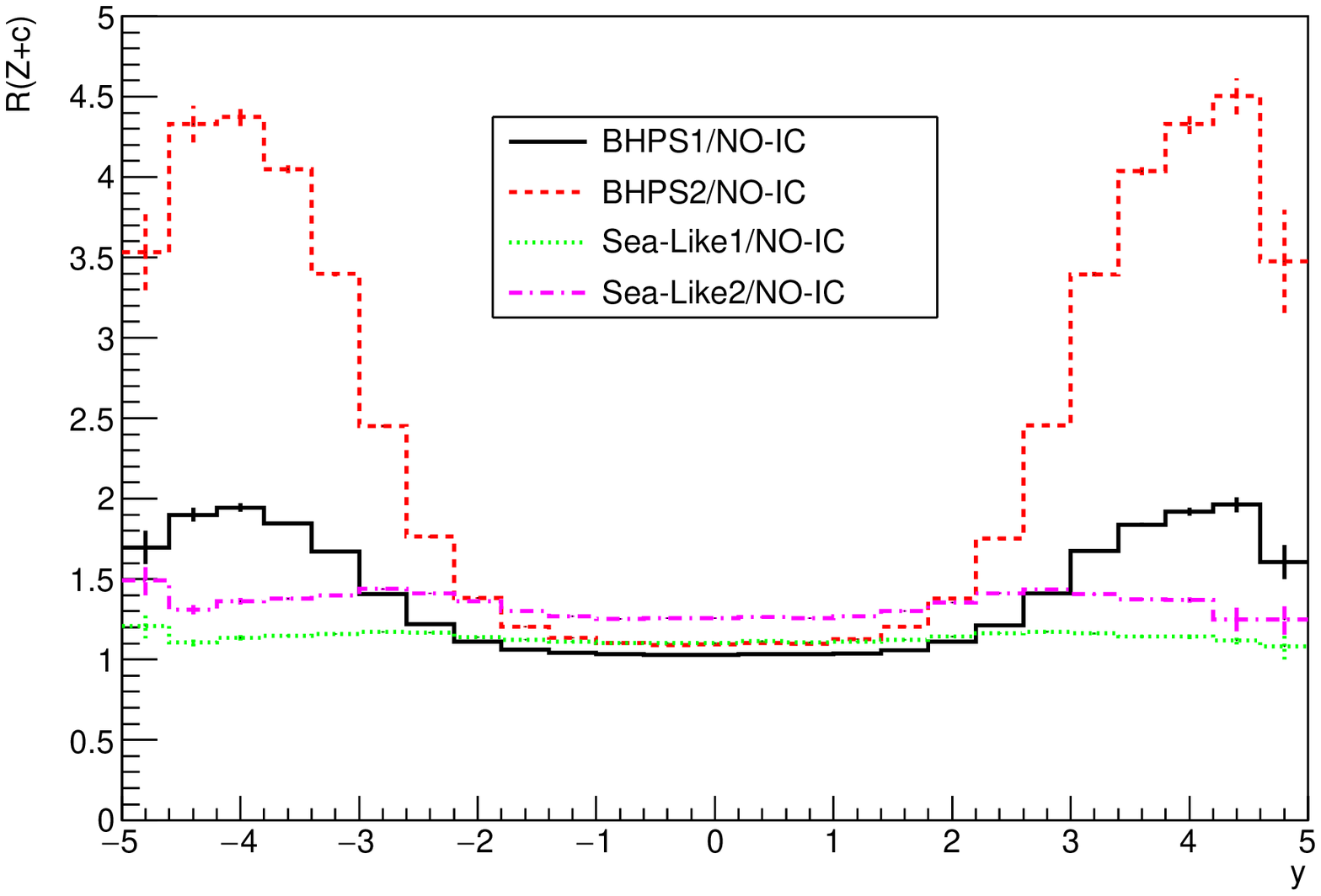,width=8cm}} \\
{\psfig{figure=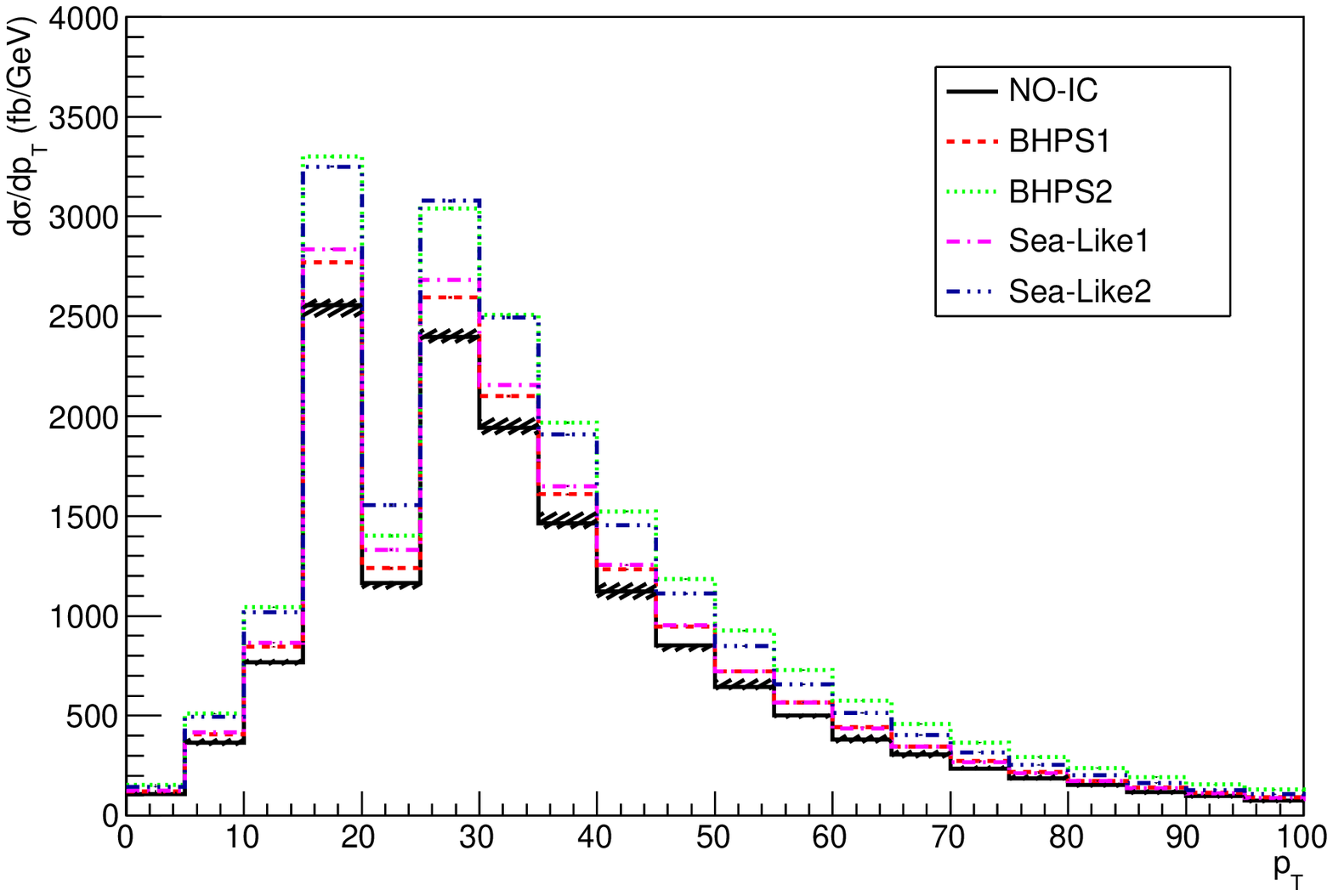,width=8cm}} & {\psfig{figure=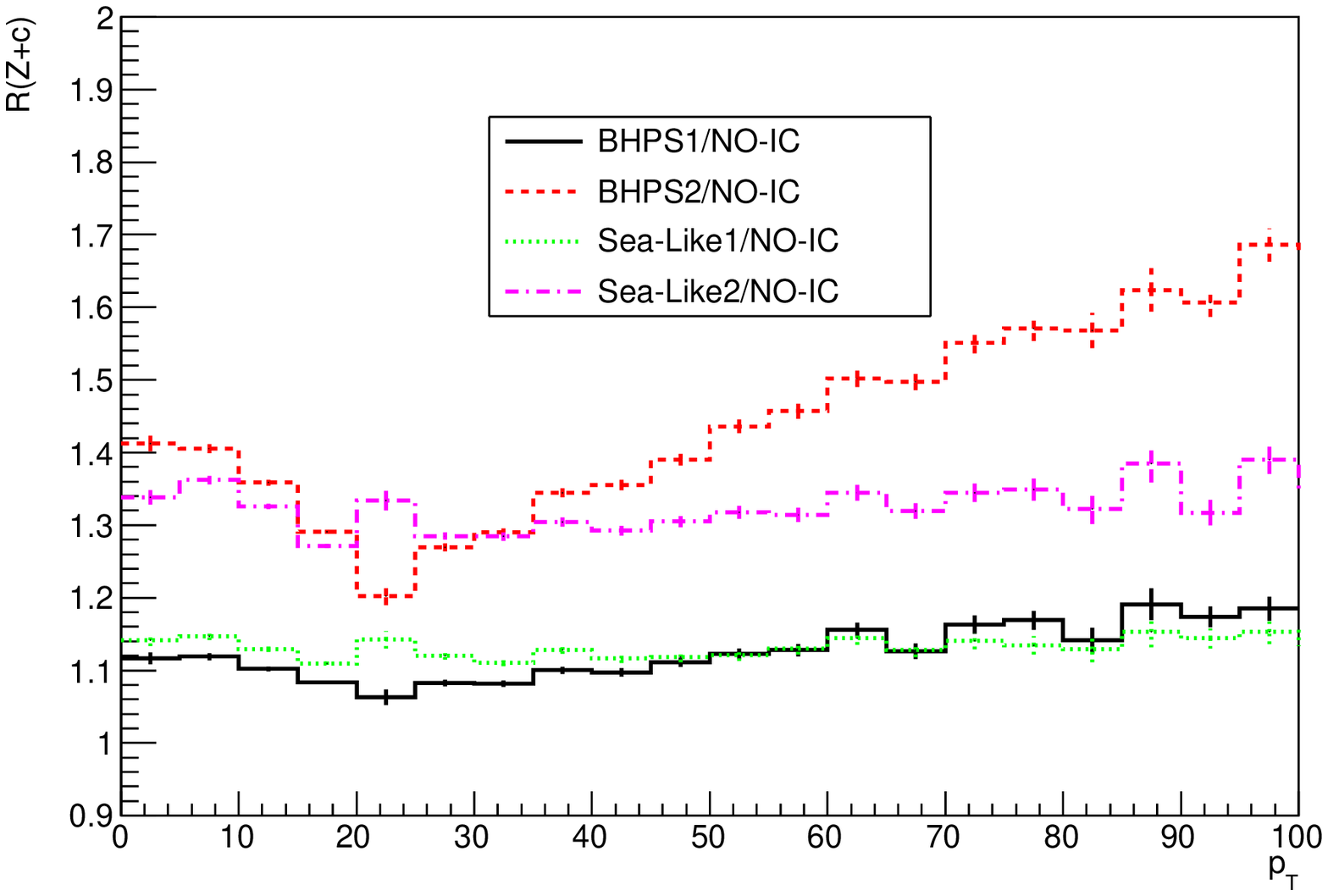,width=8cm}}
\end{tabular}                                                                                                                       
\caption{Left panels: Predictions for the rapidity and transverse momentum distributions for the $Z+c$ production  in $pp$ collisions at $\sqrt{s} = 7$ TeV. Right panels: Rapidity and transverse momentum dependencies of the ratio between the IC and central NO-IC predictions.}
\label{fig4}
\end{figure}

\begin{figure}[t]
\begin{tabular}{cc}
{\psfig{figure=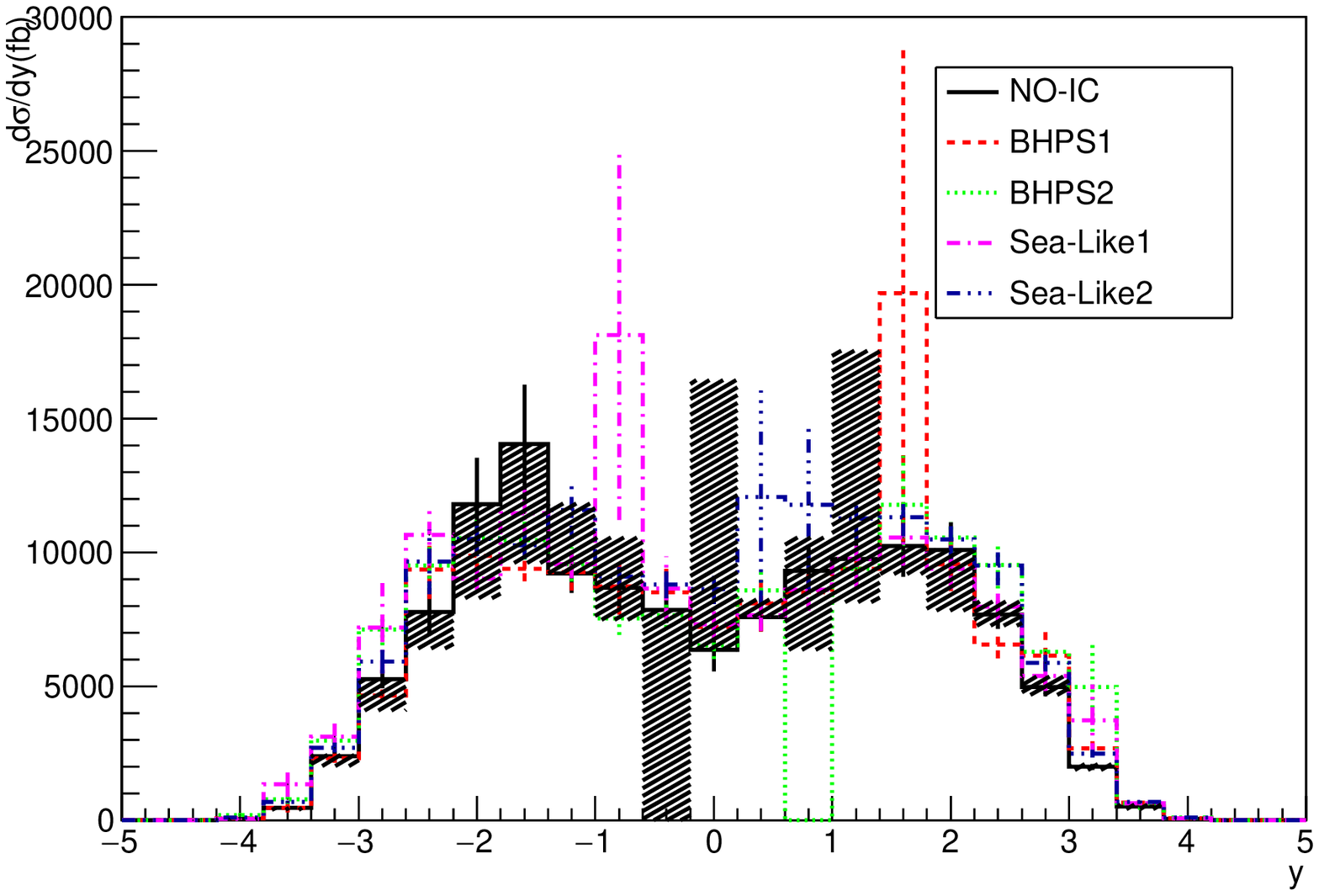,width=8cm}} & {\psfig{figure=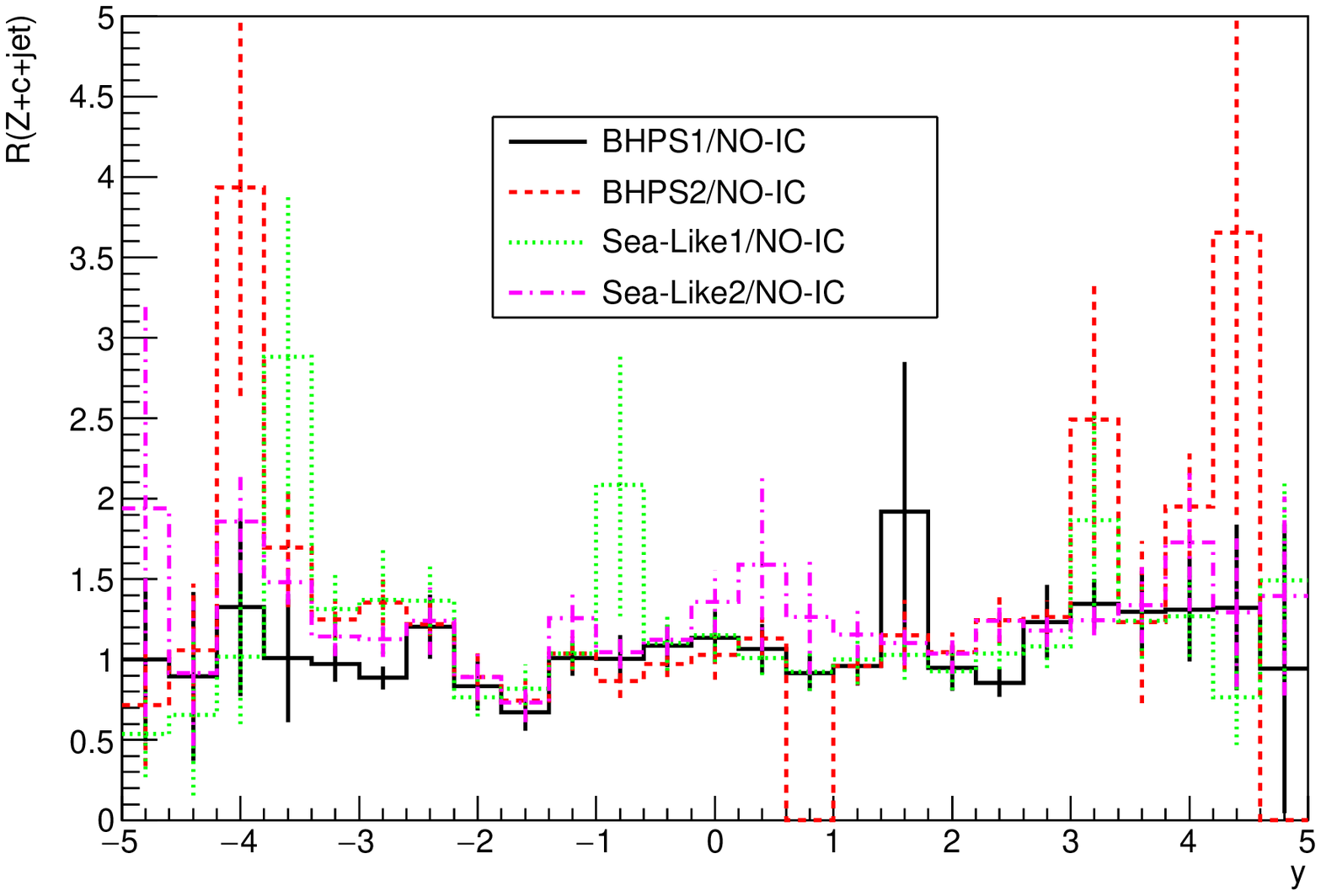,width=8cm}} \\
{\psfig{figure=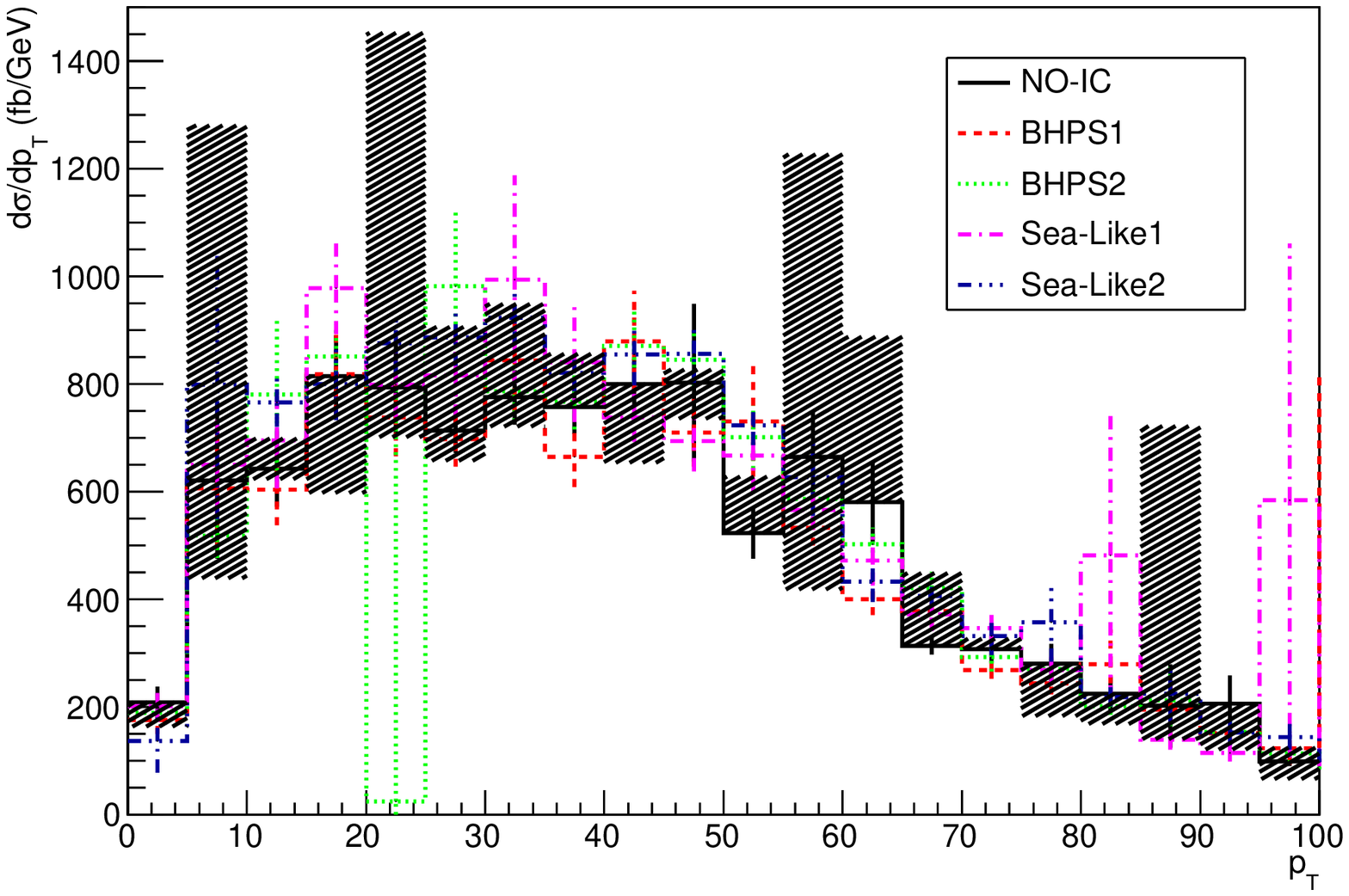,width=8cm}} & {\psfig{figure=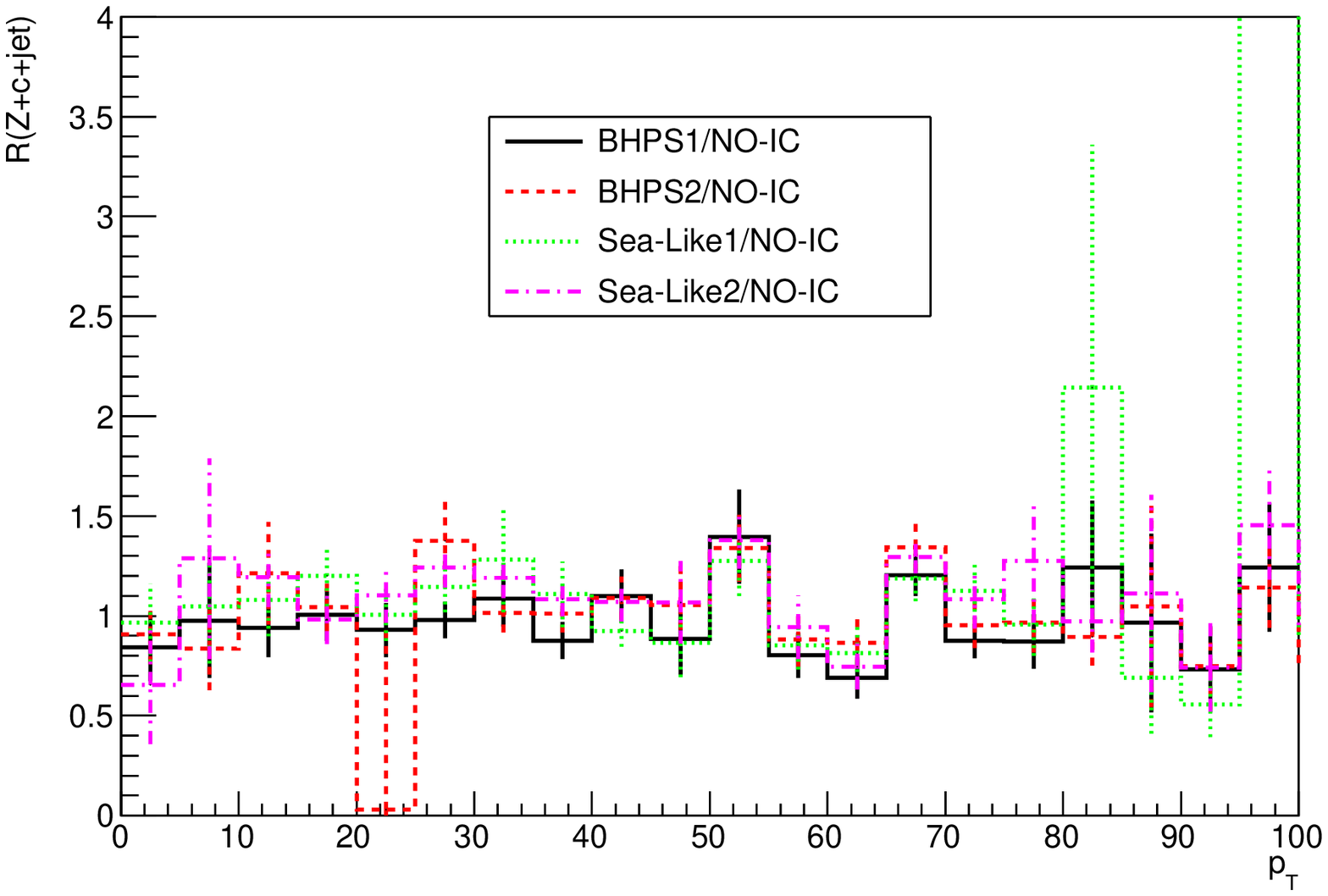,width=8cm}}
\end{tabular}                                                                                                                       
\caption{Left panels: Predictions for the rapidity and transverse momentum distributions for the $Z + c$ + jet production  in $pp$ collisions at $\sqrt{s} = 7$ TeV. Right panels: Rapidity and transverse momentum dependencies of the ratio between the IC and central NO-IC predictions.}
\label{fig5}
\end{figure}

In what follows we present our results for the rapidity and transverse momentum distributions for the $Z$, $Z+$ jet, $Z+c$  and $Z+c+$ jet cross sections considering $pp$ collisions at $\sqrt{s} = 7$ TeV and the different models for the intrinsic charm present in Ref. \cite{ct14}. Moreover, for comparison we also will present the predictions obtained using as input only the extrinsic component of the charm distribution, denoted  NO-IC in the figures, taking into account the uncertainties present in the parton distributions derived by the CTEQ - TEA group. Consequently,  our NO-IC predictions will be represented in the figures by a band of possible values for the cross sections. Finally, in order to estimate the contribution of the intrinsic charm for the process considered, we will also present the ratio between the IC and the central NO-IC predictions for the cross sections.  The analysis of the rapidity and transverse momentum distributions of this ratio allows to determine the kinematical range where  the impact of the presence of an intrinsic component in the proton wave function is larger for the  cross section.
In Fig. \ref{fig2} we present our predictions for the $Z$ production, which at leading order is proportional to $c(x_1)\bar{c}(x_2) + \bar{c}(x_1)c(x_2)$. We have that the effect of the intrinsic charm in the rapidity and transverse distribution distributions is small, as can be observed by the analysis of the ratio between IC and NO-IC predictions. Moreover, the uncertainty band in the NO-IC predictions is small. These results are expected, since the cross section is dominated by the contribution of light quarks and the uncertainty in its PDFs at large hard scales is small. A similar conclusion is derived from the analysis of our predictions for the $Z+$ jet production presented in Fig. \ref{fig3}. However,  in this case the uncertainty band is larger due to the contribution of the gluon initiated subprocesses. The small impact of the intrinsic charm in the $Z$ and  $Z+$ jet production is an important aspect, which will be explored in the analysis of the ratio between the cross sections of different processes to be discussed below. In Fig. \ref{fig4} we present our predictions for the $Z + c$ cross section \cite{camp_zc}, which for leading order is proportional to the partonic subprocess $g + c \rightarrow c + Z$ and consequently directly dependent on the charm content of the proton. We have that the uncertainty band in the NO-IC predictions is in general small, with the  IC predictions being above the uncertainty.
In agreement with previous studies  \cite{lik_prd}, we obtain that the behaviour of the transverse momentum distribution  at large $p_T$ 
is modified by the presence of an intrinsic component in the charm distribution. In particular, the BHPS2 model implies an enhancement of a factor 1.7 at $p_T = 100$ GeV. In the case of the rapidity distribution, the intrinsic charm implies a large {enhancement}  at forward rapidities, with the BHPS1 and BHPS2 predictions implying an increasing of factor 2 and 4.5 for $y = 4$. It is important to emphasize that such enhancements can be probed by the analysis of this process in the LHCb detector.   
Finally, in Fig. \ref{fig5} we present our predictions for the $Z + c +$ jet cross section \cite{camp_zcjet}. In this case the predictions are not strongly affected by an intrinsic charm component, which is associated to the fact the diagrams where the charm is not present in the initial state contribute significantly for this process. Moreover, we have that this process is strongly affected by the uncertainty present in the NO-IC PDFs.

\begin{figure}
\begin{tabular}{cc}
{\psfig{figure=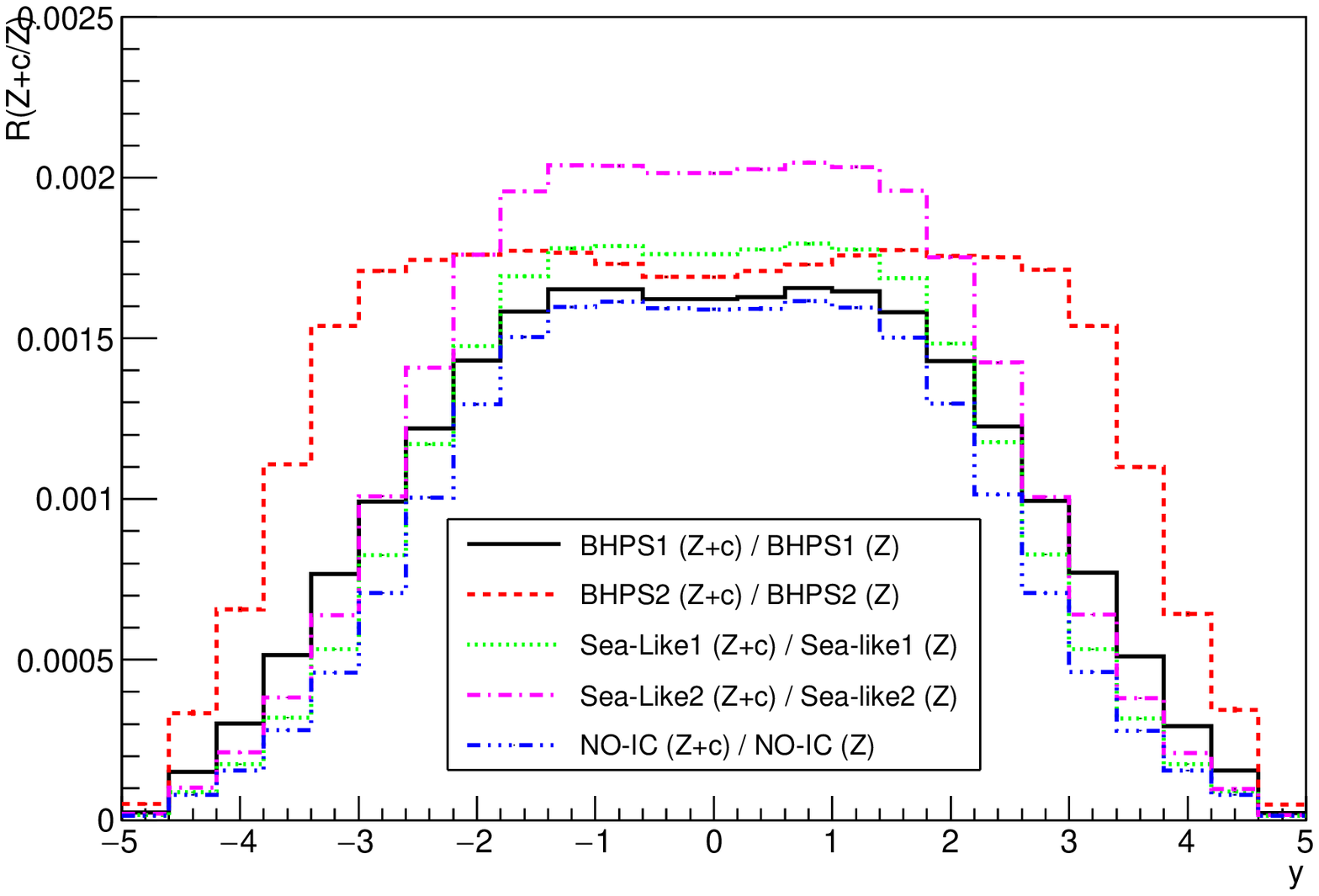,width=8cm}} & {\psfig{figure=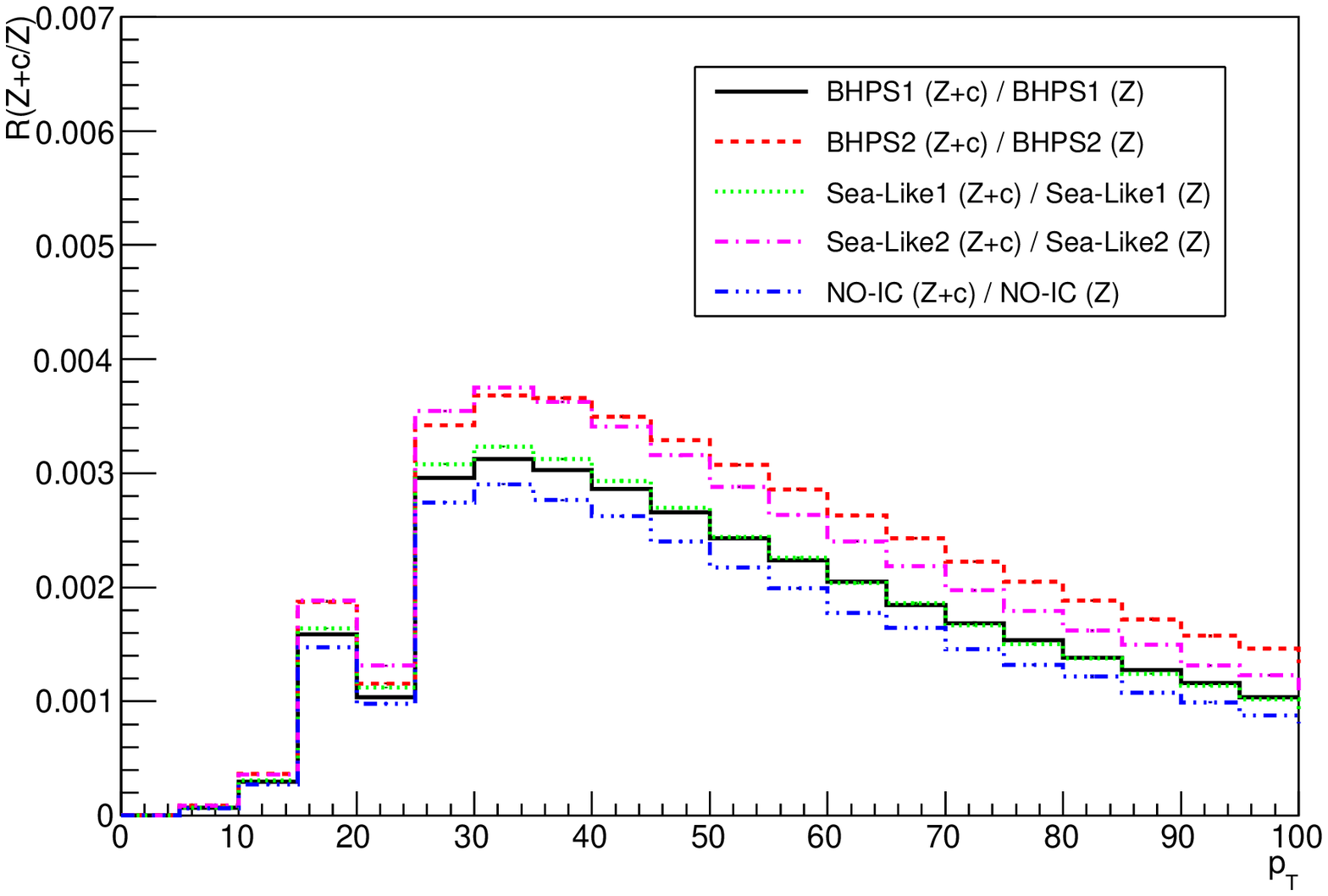,width=8cm}} \\
{\psfig{figure=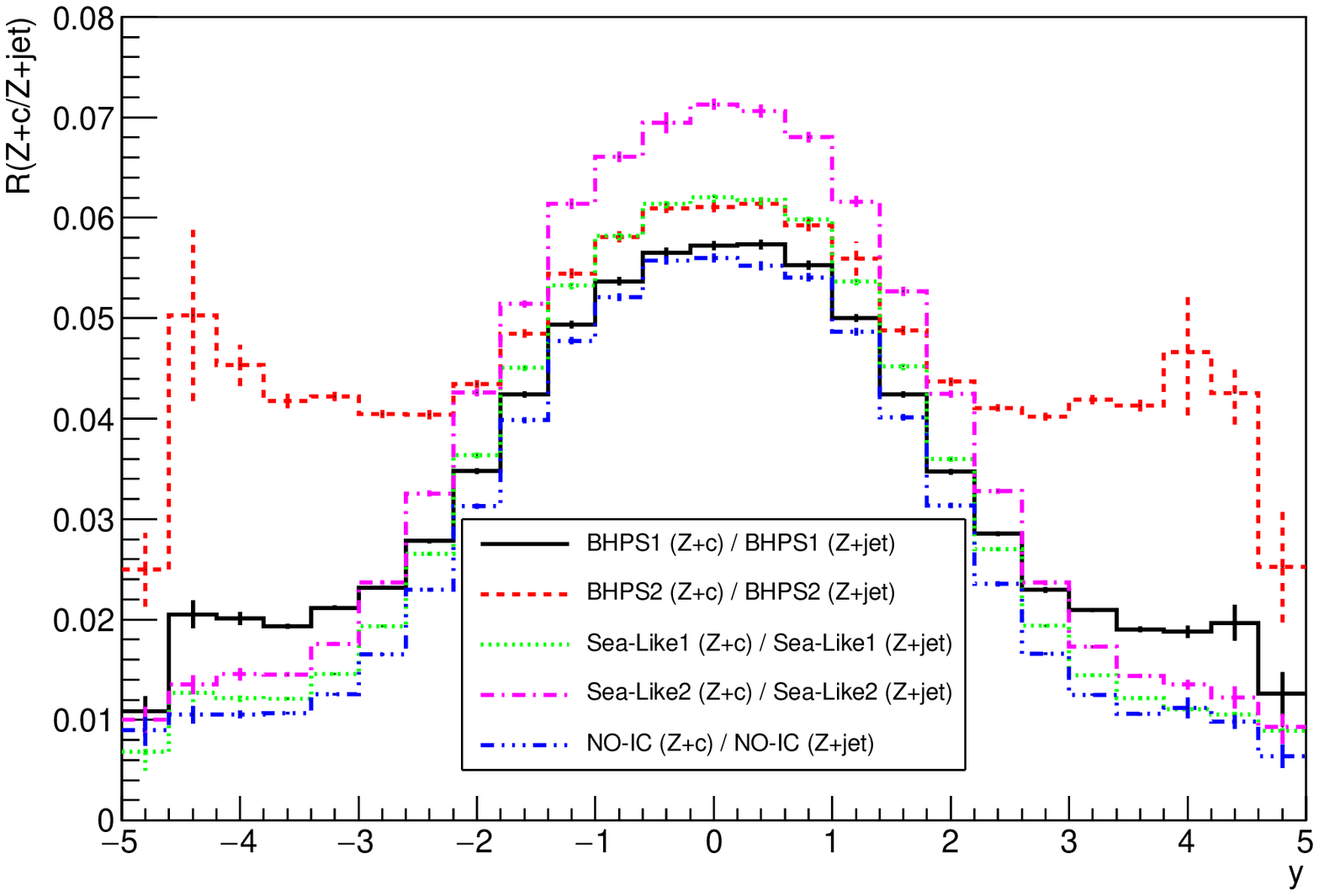,width=8cm}} & {\psfig{figure=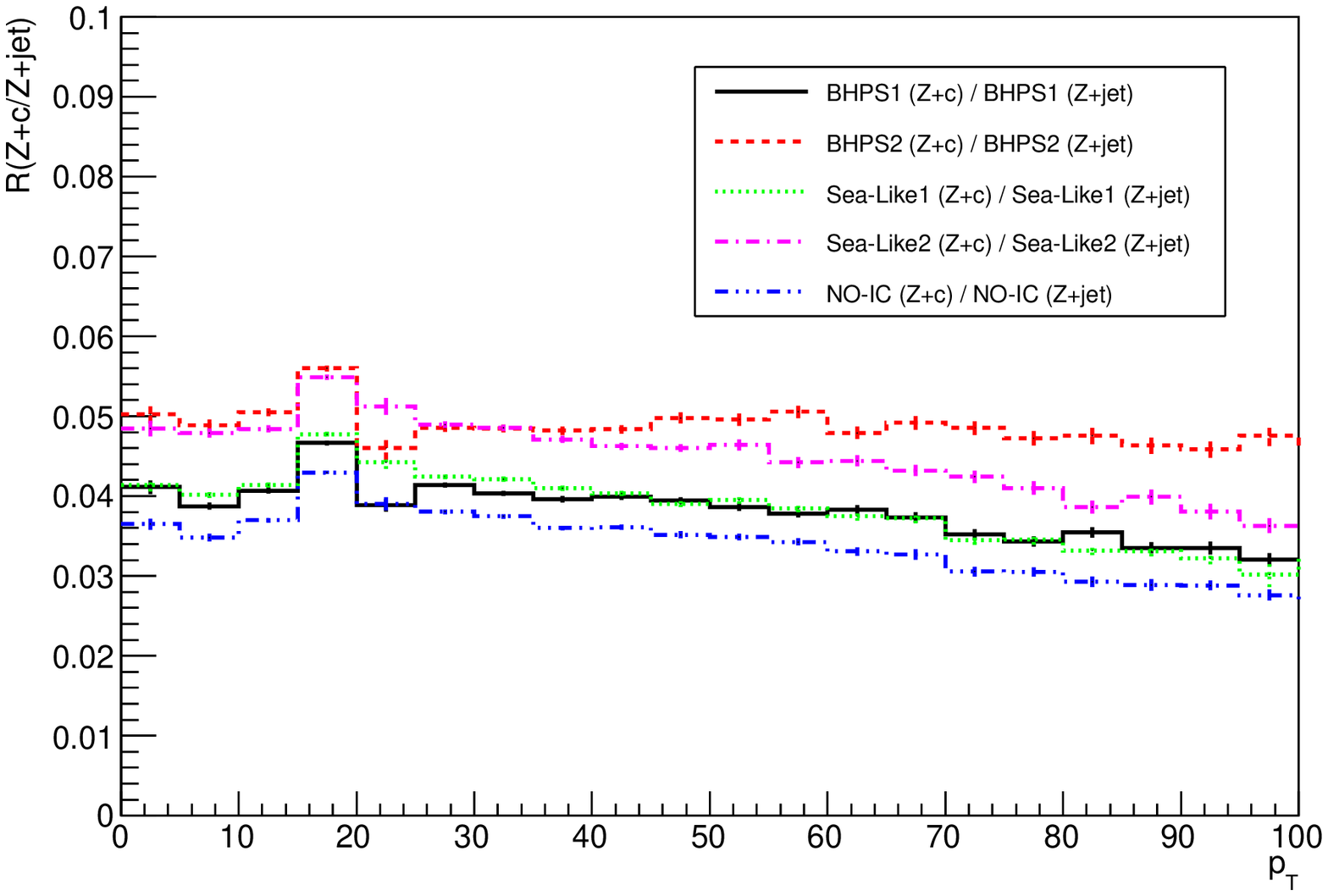,width=8cm}}
\end{tabular}                                                                                                                       
\caption{Rapidity  and transverse momentum dependencies of the ratios (a) $R(Z+c/Z) \equiv \sigma(Z+c)/\sigma(Z)$ (upper panels) and (b) $R(Z+c/Z) \equiv \sigma(Z+c)/\sigma(Z+\mbox{jet})$ (lower panels). }
\label{fig6}
\end{figure}

The previous results indicate that the ideal process to probe the existence of an intrinsic charm 
is the $Z+c$ production at large transverse momentum and forward rapidities, which is the kinematical range where the large-$x$ behaviour of the charm PDF of the proton is probed. However, as the $Z+c$ cross section is sensitive to experimental  and theoretical uncertainties, it is interesting to analyse other associated observables where these uncertainties are reduced. Here we propose the analysis of the rapidity and transverse momentum dependencies of the  ratios  $R(Z+c/Z) \equiv \sigma(Z+c)/\sigma(Z)$ and   $R(Z+c/Z+\mbox{jet}) \equiv \sigma(Z+c)/\sigma(Z+\mbox{jet})$.
An advantage of the analysis of these ratios is that PDFs uncertainties are strongly reduced, since numerator and denominator should be estimated using the same PDF set in order to be theoretically consistent.  As demonstrated before, the $Z$ and $Z+\mbox{jet}$ cross sections are almost independent of the intrinsic charm, which implies that these ratios are directly dependent on the effects present in the $Z+c$ process.
In Fig. \ref{fig6} we present our predictions for these ratios, where we have estimated the different cross sections considering a common input for the PDF's. The baseline for comparison is the prediction obtained using the NO-IC PDF's as input in the calculations. In the case of the ratio $R(Z+c/Z)$, we obtain that the Sea-like models for the  intrinsic charm implies that the ratio is enhanced by 10 -- 20\% at central rapidities, while the BHPS one implies an enhancement between 50\% and 300\% for $y = 3$. Moreover, we obtain that the intrinsic charm  implies an enhancement in the $p_T$ distribution of the ratio. Finally, for the ratio  
$R(Z+c/Z+\mbox{jet})$ we obtain a larger impact of the intrinsic charm, with analysis of this ratio at central (forward) rapidities being an important probe of the Sea-like (BHPS) models.

Lets summarize our main conclusions. Although the direct measurements of heavy flavors in DIS and hadronic colliders are consistent with a perturbative  origin, these experiments are not sensitive to heavy quarks at large $x$. Therefore, it is  fundamental to study other observables which may be used to determine the presence (or not)  of an intrinsic heavy quark component in the hadron wave function.
In recent years, a series of studies have discussed in detail the 
probe of this intrinsic component, with particular emphasis in processes that are strongly sensitive to the charm in the initial state. Our goal in this paper was to contribute for this theoretical effort by the analysis of $Z$ - boson production and related processes in $pp$ collisions at the LHC. In particular, we have discussed the $Z$, $Z+$ jet, $Z+c$  and $Z+c+$ jet cross sections considering  different models for the intrinsic charm. Our results indicated that differently from the other processes, the $Z+c$ cross section is strongly affected by the presence of the intrinsic charm. Moreover, we proposed the analysis of the ratios $R(Z+c/Z)$ and   $R(Z+c/Z+\mbox{jet})$ and demonstrated that these observables can be used as a probe of the intrinsic charm. Although we have presented results only for $\sqrt{s} = 7$ TeV, similar effects are expected for larger energies. Finally, our results indicated that the analysis of these processes at large transverse momentum and 
forward rapidities is  fundamental  to test the intrinsic charm hypothesis, which makes the LHCb experiment the ideal  laboratory. This conclusion motivates a dedicated analysis considering the experimental characteristics of the LHCb detector, which we postponed for a future publication. 

\begin{acknowledgments}
We would like to thanks to M. S. Rangel for the helpful discussions in the earlier stages of this project. G. Bailas thanks  G. G. da Silveira for the support in the numerical implementation of this study.  
This work was  partially financed by the Brazilian funding agencies CNPq, CAPES and FAPERGS.

\end{acknowledgments}

\hspace{1.0cm}

\end{document}